\documentclass[article,aps,prb]{revtex4}

\usepackage{amsmath}
\usepackage{amssymb}
\usepackage{graphicx}
\usepackage{epsfig}
\usepackage{bm}
\usepackage{units}
\usepackage{subfigure}
\usepackage{hyperref}
\usepackage{breakurl}

\newcommand{\be}{\begin{equation}}
\newcommand{\ee}{\end{equation}}
\newcommand{\bea}{\begin{eqnarray}}
\newcommand{\eea}{\end{eqnarray}}
\newcommand{\ben}{\begin{enumerate}}
\newcommand{\een}{\end{enumerate}}
\newcommand{\bit}{\begin{itemize}}
\newcommand{\eit}{\end{itemize}}

\newcommand{\bert}{\raise-0.45mm\hbox{\Large$\Box$}}

\begin{document}

\title{Towards a microeconomic theory of the finance-driven business cycle}
\author{Alejandro Jenkins}\email{ajenkins@fisica.ucr.ac.cr}
\affiliation{Escuela de F\'isica, Universidad de Costa Rica, 11501--2060 San Jos\'e, Costa Rica}
\affiliation{Academia Nacional de Ciencias, 1367--2050, San Jos\'e, Costa Rica}
\date{First version 2 Dec.\ 2013; last revision 30 Jun.\ 2015.  Published in {\it Laissez-Faire} 42, 12--20 (2015).}

\begin{abstract}

I sketch a program for a microeconomic theory of the main component of the business cycle as a recurring disequilibrium, driven by incompleteness of the financial market and by information asymmetries between borrowers and lenders.  This proposal seeks to incorporate five distinct but connected processes that have been discussed at varying lengths in the literature: the leverage cycle, financial panic, debt deflation, debt overhang, and deleveraging of households.  In the wake of the 2007--08 financial crisis, policy responses by central banks have addressed only financial panic and debt deflation.  Debt overhang and the slowness of household deleveraging account for the Keynesian ``excessive saving'' seen in recessions, which raises questions about the suitability of the standard Keynesian remedies. \\



\end{abstract}

\maketitle


\section{Introduction}
\label{sec:intro}

The business cycle is the recurrence of large macroeconomic fluctuations about the long-term growth trend in a market economy.  Understanding this phenomenon is a central and longstanding problem in economic theory. It is also a question of great political significance in the modern world.

Finance fulfills the useful function of transferring the control of capital from those who own it to those who can put it to productive use.  The vastly higher standard of living of industrialized countries compared to more traditional societies would be impossible without private finance.  There is, however, a widespread and longstanding sense that the most severe fluctuations in free market economies are driven by an {\it instability} associated with financial speculation.\cite{Weber-crises}

The evidence of business cycles is clear, but the phenomenon has proved challenging to comprehend theoretically.  One reason is that, if applicable, the microeconomic equilibrium theorems would rule out the kind of mis-coordination between economic agents that seems to mark the most severe crises. Those theorems assume that markets are complete (so that all possible voluntary transactions can be carried out) and that all agents enjoy perfect information.\cite{Arrow-Debreu}  Those conditions are obviously not achieved in practice, but markets usually equilibrate quite well under more realistic circumstances, both in the real world and in controlled experiments.\cite{Smith-rationality}

Some macroeconomic fluctuations are undoubtedly caused by the market's response to exogenous shocks.\cite{Lucas}  Credit restrictions can amplify the effects of such shocks,\cite{Kiyotaki} but even then the welfare cost of the fluctuations might be insufficient to justify interventionist stabilization policies.\cite{Krusell}  On the other hand, the largest of the macroeconomic fluctuations, associated with episodes of financial panic followed by high unemployment and political disturbances, seem to be {\it endogenous} and {\it inefficient}.  In order to understand what could drive such fluctuations, and how to alleviate them, it is necessary to understand just why and how financial markets may sometimes fail to equilibrate.

Here I shall sketch, in broad outline, a program for a {\it microeconomic} theory of the business cycle, i.e., for an understanding based on the choices made by rational agents.  In this view, the business cycle, or at least a major component of it, appears as a recurring market inefficiency driven by incompleteness of financial markets and by information asymmetries between borrowers and lenders.

This proposal brings together five distinguishable but connected processes.  Previous researchers have treated each of these, though, to my knowledge, they have not articulated them as a coherent theoretical framework for the business cycle as a whole. These processes are: $(1)$ the leverage cycle, $(2)$ financial panic, $(3)$ debt deflation, $(4)$ debt overhang, and $(5)$ deleveraging of households.

The application to the business cycle of ideas taken from the mathematical sciences of control theory, non-linear dynamics, and non-equilibrium thermodynamics has a long but not very fruitful history.\cite{SO}  I believe that the basic shortcoming of such efforts has been the absence of an adequate microeconomic foundation, which is what a physical scientist is least likely to contribute.  Nonetheless, I have been motivated to draw up this ``memorandum'' by the sense that the separate strands of what could be a satisfactory understanding of the finance-driven business cycle exist already, but are not being drawn together in a wholly satisfactory way by economists.  Moreover, the policy implications of such a theory could be significant, but I have not seen them clearly articulated in the public debates that have followed the financial crisis of 2007--08.

Of existing strands of macroeconomic thought, the most relevant to this work is traceable to Irving Fisher (1867--1947), the first theorist to stress that debt need not be macroeconomically neutral.\cite{Tobin}  Hyman Minsky has pursued this idea within a post-Keynesian framework.\cite{Minsky}  More recently, Claudio Borio has proposed a ``financial cycle'' as the driver of the biggest macroeconomic fluctuations.\cite{Borio}  Steven Gjerstad and Vernon Smith (who have had the most direct influence on my own thinking) have explored related ideas, with a particular focus on residential mortgages.\cite{Rethinking}

\section{Leverage cycle}
\label{sec:leverage}

It is widely recognized that asset markets (in which the same good may be purchased and sold many times) are susceptible to speculative bubbles, in which prices rise well above the fundamentals because of a self-reinforcing but finally unsustainable expectation that prices will continue to rise.  This process ends with a rapid collapse of the asset price.  Despite the widespread recognition that this phenomenon has recurred throughout history, the microeconomic theory of why such bubbles occur and how they may harm the economy as a whole remains incomplete and contentious.

That self-reinforcing fads should cause temporary spikes in the prices of some assets may not require an explanation much deeper than the obvious psychological one.  Nothing in the equilibrium theorems prevents people's preferences from changing over time.  But the sheer size of certain speculative bubbles, and the serious damage that they can cause to the economy at large, pose major challenges for economic theory and policy.

According to Max Weber's {\it General Economic History}, the great Mississippi Company bubble of the late 1710's ``can be explained only by the fact that short selling was impracticable since there was as yet no systematic exchange mechanism.''\cite{Weber-short}  That is, the bubble resulted from a market incompleteness that made it easier for optimists than for pessimists to bet on future price changes.

Modernly, the relevant incompleteness probably lies in the financial market rather than the asset market itself: it is usually easier for optimists than for pessimists to leverage their bets (i.e., to make them with borrowed money).  As the price of an asset begins to climb, it may happen that it can be increasingly leveraged.  As leverage increases, it takes a smaller up-front payment to acquire the asset.  The marginal buyer, who sets the price, is therefore likely to be more optimistic about the price trend, making the price higher.  Higher prices make the asset a more attractive investment, justifying optimism and leading to greater leverage.  This positive feedback between leverage and asset price continues until they reach unsustainable levels and collapse.  John Geanakoplos has developed a sophisticated model for this process.\cite{Geanakoplos}

Geanakoplos points out that in the run-up to the 2007--08 financial crisis, credit default swaps (CDS) did provide a mechanism to borrow money for betting against the housing market.  (A CDS is a negotiable contract that promises to compensate the buyer if some underlying financial instrument goes into default.)  However, CDS's became widely available for US residential mortgages only in late 2005, when the housing bubble was already well underway and approaching its peak.  In those circumstances, the introduction of CDS's may have helped precipitate the crash.\cite{CDS}

A point worth highlighting is that the standard equilibrium theorems allow for differences in individual preferences, but not for differences of beliefs (e.g., of optimists versus pessimists), which would not exist in a state of perfect knowledge.  Hayek,\cite{Hayek} Stigler,\cite{Stigler} and other pioneers of the economics of information have emphasized that, in the real world, economically relevant knowledge is not a given, but rather emerges through the market process itself.\cite{Smith-info}  Therefore, the absence of CDS's in the housing market may have been significant not only as an incompleteness {\it per se}, but also because it prevented the information possessed by the more pessimistic potential investors from being incorporated into the asset prices in a timely way.

\section{Financial panic}
\label{sec:panic}

When the growth phase of the leverage cycle ends with a rapid decline of asset prices, many of those who borrowed to purchase the asset are forced to default on their loans.  Since the now depressed asset usually serves as collateral for those loans, lenders suffer serious losses.  Borrowers whose debts are now worth more than the collateral are especially likely to default.  Furthermore, financial institutions often hold those same assets in their balance sheets.  They are therefore suddenly faced with large accounting short-falls and acute uncertainty about their ability to meet their obligations.  A panic ensues in which other lenders (e.g., ordinary bank depositors) withdraw their funds and financial institutions are forced to liquidate assets at ``fire sale'' prices, further depressing their market value.  Many financial institutions may become illiquid and the financial sector as a whole may cease to function properly, doing immediate harm to the real economy.\cite{Smith-bubble}

Such a financial panic may be either alleviated or aggravated by financial regulations and government interventions.  Since the early 20th century, it has been widely accepted that one of the key roles of central banks is to act as lender of last resort, providing liquidity to distressed financial institutions that are actually solvent.\cite{Bernanke-panic}  Nonetheless, some questions remain about how best to implement this in practice.  It may be difficult for the lender of last resort, in the midst of a panic, to make fully rational decisions about which distressed institutions are merely illiquid (so that loans are likely to be repaid when conditions normalize) and which institutions are actually insolvent (in which case efforts to keep them from facing bankruptcy may hinder the financial sector's return to health and drag out the consequences of the bursting of the leveraged asset bubble).\cite{Smith-balance}  The bailing out of financial institutions with taxpayer money may also create a ``moral hazard'', encouraging others to make high-risk speculations in the future.

\section{Debt deflation}
\label{sec:deflation}

Irving Fisher, who famously failed to see the stock market crash of 1929 coming, developed a deep understanding of the role of finance and money in the ensuing Great Depression.  His principal contribution in this area was the theory of debt deflation.\cite{deflation}  In the modern financial system, money can have a ``perverse elasticity'', i.e., its supply can shrink when the demand increases.  The underlying reason for this is that the currency actually issued by the central bank (the monetary base) is only a fraction of the total money supply.  The rest of it is generated by commercial banks in the process of lending out part of their demand deposits.  When a financial panic strikes, demand for liquidity rises across the board and lending by commercial banks contracts, causing the supply of money to drop.  The value of money may therefore rise sharply, leading to deflation.

Unanticipated deflation makes it more difficult for borrowers to meet their obligations to lenders.  This worsens the debt overhang situation, which we will discuss in the next section.  Thanks to the work of Milton Friedman and other ``monetarists'' who followed in Fisher's footsteps, avoiding deflation during economic downturns is now widely recognized as a priority for central banks.  Since modern central banks can issue currency at will, it seems possible in general to avoid deflation by a sufficiently aggressive intervention, though political considerations and concerns over long-term price stability can make this hard to carry out in some cases.\cite{Bernanke-deflation}

\section{Debt overhang}
\label{sec:overhang}

It has long been recognized that indebtedness may, in some circumstances, reach such high levels that overall economic output is negatively affected.  This may also breed social unrest and political pressure for government-mandated debt relief.\cite{Weber-Greece}  Clearly, an important factor in this is the information asymmetry between borrower and lender, which can lead to serious agent-principal problems.\cite{principal-agent}  In particular, when a borrower's net worth is negative (i.e., when his debts are worth more than his assets, a situation referred to as being ``underwater''), the bulk of the borrower's income must go directly to the lender.  The borrower therefore has a reduced incentive to undertake profitable projects.  Lenders will recognize this and become unwilling to extend further credit, even when it could help the borrower to improve his financial situation and repay his debts.

This situation, commonly called ``debt overhang'', is avoided in the normal course of modern private finance because it is in the interest of neither borrowers nor lenders.  It may, however, emerge suddenly as a consequence of the leverage cycle.  After the bursting of a finance-driven asset bubble, the net borrowers (henceforth, for simplicity, ``households'') will be left as owners of the depressed assets and the net lenders (henceforth simply ``banks'') will be left as owners of the debt.  Once the fog of the financial panic has cleared, a new landscape is revealed in which a large amount of wealth has passed from households to banks.  This may leave many households underwater and unable to spend or invest at normal levels.

Under such conditions, banks will have little incentive to extend further credit to households, which makes it more difficult for households to dig themselves out from under their load of debt.  Banks will instead direct new investment towards the very safest assets, such as government bonds and gold (``flight to quality'').  The situation may be exacerbated by debt deflation (if the central bank does not combat it successfully) since it induces a further transfer of wealth from households to banks.  This ``financial accelerator'' effect seems to account for some of the peculiar dynamics of the economy after a financial panic.\cite{accelerator}

A question of both theoretical and practical importance is why banks do not voluntarily condone some of the debts of households that are underwater.  It may simply be that a bank usually lacks the information to make rational decisions about which debts to forgive in order to improve its own revenue stream.  (This is somewhat akin to the point raised earlier about the difficulty for the lender of last resort in discriminating between merely illiquid institutions and those that are actually insolvent.)  Uncertainties about the government's regulatory, monetary, and fiscal response to the crisis may make it even more difficult for individual banks to pursue rational policies of debt relief or renegotiation.

\section{Deleveraging}
\label{sec:deleveraging}

Empirically, it is clear that conditions of depressed aggregate demand may continue long after the financial panic phase of the cycle is over.  Keynes merely asserted the reality of such episodes of persistent disequilibrium, in which ``excessive saving'' is not cured by low interest rates.  He then invoked the rigidity of wages and other prices in order to explain why it could cause chronic unemployment.\cite{Keynes}

Household debt overhang may provide a microeconomic foundation for the otherwise mysterious fall of aggregate demand and for the fact that additional money pumped into a depressed economy tends to be hoarded rather than spent (the Keynesian ``liquidity trap'').\cite{traps}  According to the scheme that I have sought to outline here, these would be manifestations of the principal-agent problem between banks and households left underwater after the end of the leverage cycle's growth phase.  Under such conditions, additional money provided to households goes primarily towards paying down their debts, while banks are unlikely to reinvest those payments in the households as long as they remain underwater.

In other words, the principal-agent problem between the banks and the underwater households impairs the ability of the financial sector to achieve its purpose of finding efficient uses for capital.  This is manifested macroeconomically as a Keynesian episode of ``excessive saving'' and reduced aggregate demand.  This is consistent with the evidence that serious recessions often follow a large fall in {\it housing} prices,\cite{housing-cycle} since residential mortgages are the most widespread form of leveraged investment, while housing is the main asset held by ordinary households.  The slow recovery from a recession would then be directly tied to the deleveraging of underwater households.\cite{Rethinking}

\section{Implications for policy}
\label{sec:policy}

The monetary component seems to be the aspect of the business cycle that is best understood by both academic economists and policy makers.\cite{Austrian}  Aggressive efforts by the US Federal Reserve to combat deflation probably helped prevent the Great Recession of 2007--08 from achieving destructive proportions equal to or greater than those of the Great Depression of 1929.  The policy of ``quantitative easing'' (under which a large amount of new money was created and used to purchase government bonds from the banks) can be seen as having met the banks' greatly elevated demand for liquidity and will probably end up generating a net profit for the Federal Reserve.\cite{QE}  Monetary policy, however, can only address the problem of debt deflation and it leaves much of the dynamics of the business cycle untouched.

Perhaps the most problematic element of the scheme outlined here, both theoretically and in terms of policy, is the growth phase of the leverage cycle.  Many of the greatest economists of their day (Irving Fisher and Ben Bernanke, to name just two) conspicuously failed to see the asset price bubbles that would soon burst and trigger a major economic crisis.  Indeed, as Bernanke has since pointed out,\cite{Bernanke-MM} the Modigliani-Miller theorem on the irrelevance of the capital structure of firms seemed to rule out the possibility of the financial system having such significant and destructive macroeconomic consequences.\cite{MM}  Geanakoplos's theory of the leverage cycle does not offer any clear criterion for when the rate of leverage has become dangerous.

If the leverage cycle is indeed the principal mechanism behind asset price bubbles, it is important to understand just how it evades the equilibrium theorems.  If the problem lies chiefly with the market incompleteness that makes it more difficult for pessimists than for optimists to leverage their bets on asset prices, then it might not require much further regulation or direct government intervention to keep the amplitude of the leverage cycle in check.  Perhaps it would suffice for instruments such as CDS's to become widely available.  If this were insufficient, it might be necessary to mandate minimum margin requirements for mortgage markets, like the ones that have long applied in stock exchanges.\cite{Smith-balance}

If the microdynamics of the fall of aggregate demand and of the liquidity trap associated with deep recessions are indeed tied to households being underwater, then the usual Keynesian remedies for recessions appear problematic.  Deficit spending by the government (which transfers some of the household debt to the public sector) and inflation (which reduces the real value of the outstanding debt) can ease deleveraging and therefore alleviate the recession.\cite{Keynesians}  However, neither policy can be focused on the problem of underwater households.  Moreover, they can have major undesirable consequences, particularly in the way in which they are likely to be pursued by elected governments.\cite{Buchanan}

\section{Conclusions}
\label{sec:conclusions}

Financial markets normally help the approach to an efficient equilibrium, a condition in which every agent would enjoy the greatest satisfaction possible without making another worse off.  Here I have outlined a possible theory of the business cycle that, if successful, would explain how and why those same financial markets sometimes exhibit instabilities that have the opposite effect, driving the economy into recurring episodes of disequilibrium.

The long-term growth and general resilience of free market economies, compared to other forms of economic organization, suggest that the gains from financial speculation have substantially outweighed the pain of the finance-driven business cycle.\cite{Shiller}  But the recurrence of financial crises has long been a political liability of the free market system.  Nearly a century ago (and well before the Great Depression), Max Weber taught his students of economic history that:

\begin{quote}
Crises in the broader sense of chronic unemployment, destitution, glutting of the market and political disturbances which destroy all industrial life, have existed always and everywhere. But there is great difference between the fact that a Chinese or Japanese peasant is hungry and knows the while that the Deity is unfavorable to him or the spirits are disturbed and consequently nature does not give rain or sunshine at the right time, and the fact that the social order itself may be held responsible for the crisis, even to the poorest laborer.  In the first case, men turn to religion; in the second, the work of men is held at fault and the laboring man draws the conclusion that it must be changed.\cite{Weber-quote}
\end{quote}

In the 1930s, the effects of the Great Depression greatly contributed to the political advance of socialism, which in its various forms sought to bring capital under state control.  Socialism, however, is faced with the great challenge of defining and implementing rational criteria for the allocation of scarce resources, in the absence of market competition.\cite{socialism}  By the late 20th century, socialism had lost much of its prestige and credibility. Improvements in the theoretical understanding of the business cycle, and therefore in the effectiveness of policy responses to it, probably helped mute the reaction against free markets in the wake of the 2007--08 crisis.

A successful theory of the business cycle would be a major scientific achievement, as well as a matter of great practical and political significance.  A plausible qualitative understanding of the relevant microeconomics seems a prerequisite for any serious effort to formulate a rigorous quantitative theory of the cycle.  The basic outline of such a qualitative understanding can, I think, finally be discerned, though much empirical and theoretical work necessarily remains to be done.


\begin{acknowledgements}

I thank Rodrigo Cubero for feedback on the manuscript and for pointing me to Ref.~\onlinecite{Borio}.  I also thank Diego Aycinena, Julio Cole, and Mark Wise for comments.  Part of this work was presented before the 1st International Congress on Actuarial Science and Quantitative Finance, in Bogot\'a, Colombia, on 20 Jun.\ 2014.

\end{acknowledgements}


\bibliographystyle{aipprocl}   

\begin{thebibliography}{99}
\bibitem{Keynesians}
Eggertsson and Krugman (see Ref.~\onlinecite{traps}) claim to justify Keynesian stimulus policies.  Their model, like that of Guerrieri and Lorenzoni, does not incorporate the dynamics of underwater households that I have sketched here.
\end{thebibliography}

\end{document}